\documentclass[aps,twocolumn]{revtex4-1}
\usepackage{graphicx,epsfig}
\usepackage{amsmath}
\usepackage[T1]{fontenc}
\usepackage{hyperref}
\hypersetup{
     colorlinks   = true,
     citecolor    = magenta,
     linkcolor    = blue,
}

\begin{document}

\title{Effects of an attractive three body interaction on a spin-1 Bose Hubbard model} \author{Sk Noor Nabi}
\author{Saurabh Basu} \email{saurabh@iitg.ernet.in}
\affiliation{Department of Physics, Indian Institute of Technology
  Guwahati, Guwahati, Assam 781039, India} \date{\today}

\begin{abstract}
We study the effects of an attractive three body interaction potential on a spin-1 ultracold Bose gas using mean field approach (MFA). For an antiferromagnetic (AF) interaction, the third MI lobe is predominantly affected, where it completely engulfs the second and the fourth MI lobes at large values of the interaction strength. Albeit no significant change is observed beyond the fourth MI lobe. The formation of the spin singlet (nematic) MI phase and the different order of phase transitions to the SF phase have been carefully scrutinized with the help of spin eigenvalues and spin nematic order parameter. In the ferromagnetic case, the phase diagram shows similar features as that of a scalar Bose gas. We have compared our results on the MFA phase diagrams for both types of the interaction potential via a perturbation expansion in both the cases.
\end{abstract}

\keywords{spinor ultra-cold atoms, extended interaction, harmonic confinement}

\maketitle

\section{Introduction}
Unlike thermodynamic phase transitions, quantum phase transitions (QPT) are governed by the quantum fluctuations and form a major backbone in exploring correlated many body phenomena in condensed matter physics. The technological advancements in cooling and trapping of the ultracold atomic gases in optical lattices have finally paved the way to observe various QPTs and hence act as a quantum simulator due to the tunability of diverse experimental parameters. 
\\\indent A prominent example of such QPT is the experimental realization of the superfluid (SF) to Mott insulator (MI) transition by magneto-optically trapped (MOT) $^{87}Rb$ atoms in an optical lattice \cite{Greiner}. The general properties of these trapped ultracold atoms in optical lattices were first outlined by Jacksh {\it{et al.}} \cite{PhysRevLett.81.3108} where one can easily go from a state of coherent superfluid (SF) to a MI phase as long as one plays with the system parameters, such as, the tunneling and the on-site interaction potential of the Hamiltonian. 
\\\indent In principle, the bosonic atoms can have multiple degrees of freedom whose occurrence entirely depend upon the trapping mechanism. In contrast to MOT which confines the atoms in a single hyperfine state thereby producing a spin-0 Bose gas, the use of sophisticated optical dipole trap helps in preserving all the hyperfine degrees of freedom and thus treat the system as a spinor Bose gas. Such an idea is now a reality after the successful observation by optically trapping $^{23}Na$ atoms where the interaction between the dipole force of neutral atoms and electric field of laser beams results in a spin-1 condensate \cite{PhysRevLett.80.2027,RevModPhys.85.1191}.

The general dynamics of spin-1 ultracold Bose gas in an optical lattice is described by an extended Bose Hubbard model which includes an additional spin dependent interaction potential due to the presence of hyperfine degrees of freedom \cite{PhysRevLett.81.742,ohmi,PhysRevA.68.063602}. Such a spin dependent interaction can be antiferromagnetic (AF) as well as ferromagnetic in nature and thus possesses a wealth of rich phase properties and quantum magnetism as a result of spontaneous symmetry breaking of the ground state. It was found that the ground state shows a $U(1)\times {\bf{S}}^{2}$ symmetry in the AF case and a $SO(3)$ symmetry in the ferromagnetic case in contrast to the usual $U(1)$ symmetry of a standard spin-0 BHM \cite{PhysRevLett.81.742,ohmi}.

To understand the origin of various phases in a spinor gas, several numerical techniques, such as mean field approximation (MFA) \cite{PhysRevA.91.043620,PhysRevB.69.094410,PhysRevB.77.014503}, quantum Monte Carlo (QMC) \cite{PhysRevA.82.063602,PhysRevB.84.064529,PhysRevB.88.104509,PhysRevLett.102.140402} and strong coupling perturbative expansion (PMFA) \cite{PhysRevA.70.043628,PhysRevLett.94.110403,PhysRevA.87.043624} have been developed to gain an in-depth knowledge on how different extent of the Mott insulating and superfluid phases will affect the ground state phase diagram. In the AF case, the MI state with even occupation densities form a spin singlet state with vanishing spin nematic order, while the odd occupation densities show a spin nematic state with a finite nematic order parameter. This phenomenon of spin singlet (nematic) formation is now responsible for an even (odd) asymmetry in the MI phase and hence shows a first (second) order transition to the SF phase \cite{PhysRevA.91.043620,PhysRevB.69.094410,PhysRevB.77.014503,PhysRevA.82.063602,PhysRevB.84.064529,PhysRevB.88.104509,
PhysRevLett.102.140402,PhysRevA.70.043628,PhysRevLett.94.110403,PhysRevA.87.043624}. However density matrix renormalization group (DMRG) studies indicate a signature of a dimerized phase and shows odd-even asymmetry for higher values of the AF interaction \cite{PhysRevLett.95.240404}.

Besides, the use of hyperfine degrees of freedom as a short lattice dimension, known as synthetic dimension helps in realizing high synthetic magnetic field \cite{PhysRevLett.117.125301} and various density ordered SF phases in presence of spin-orbit coupling (SOC) \cite{PhysRevA.91.023608,PhysRevB.93.081101}. Later, the inclusion of long range extended interactions such as density-density \cite{NOORANDP} and spin-spin \cite{PhysRevB.92.054506} extended interactions displays a charge (CDW) and spin density wave (SDW) pattern in the MI phase in the former case, while an antiferromagnetic SF (AFSF) which has both the crystalline and off-diagonal order in the latter. Moreover, the effects of magnetic fields \cite{PhysRevA.94.063613,PhysRevLett.84.1066,PhysRevLett.87.080401,Zhang,PhysRevB.70.184434,NoorEPL,PhysRevLett.112.043001}, disorder \cite{PhysRevA.83.013605,NoorJPB,PhysRevB.95.235128}, dipolar interaction \cite{PhysRevLett.97.020401,PhysRevLett.97.130404} and spin dynamics etc have now been studied extensively in the context of a spin-1 Bose gas.

Apart from studying such properties in presence of two body interactions cited above, recently, the consequences of higher body interaction, such as a repulsive three body interaction on the ground state phase diagrams have been explored on a spin-1 ultracold Bose gas using MFA in chapter 6 and DMRG \cite{PhysRevA.94.033623,arXiv:1707.08195} techniques. Interestingly, the mean field phase diagram still shows a spin nematic-singlet formation and hence an asymmetry in the MI phase in presence of both the two and three body interactions, while for a purely three body interaction such asymmetry is destroyed \cite{NoorEPL2}. On the contrary, the DMRG studies show that there is neither any asymmetry, nor any spin singlet-nematic formation in the MI phase and there is a possible phase transition involved in the SF phase in presence of both the two and three body interactions \cite{PhysRevA.94.033623,arXiv:1707.08195}. 

Besides the presence of such repulsive three body interaction, another form of three body interaction which is attractive in nature is proposed by Safavi-Naini {\it{et al.}} \cite{PhysRevLett.109.135302}. Such interaction can be observed by exciting a triple occupied state into an excited hyperfine state so that they can form a trimer state. Such type of attractive interaction was studied in the context of a spin-0 Bose gas where it is found to affect a particular MI lobe. Motivated by such interesting phenomena, it is worthwhile to explore the effects of an attractive multi-body (three body) interaction in a spinor Bose gas. We shall study the effects of an attractive three body interaction on the SBHM to see the consequences on the phase diagrams, particularly the existence of odd-even asymmetry in the MI lobes. 
\\\indent Besides the presence of such repulsive three body interaction, another form of three body interaction which is attractive in nature is proposed by Safavi-Naini {\it{et al.}} \cite{PhysRevLett.109.135302}. Such interaction can be observed by execiting a triple occupied state into an excited hyperfine state which can form a trimer state and was studied in the context of a spin-0 Bose gas where it is found to affect a particular MI lobe. Motivated by such interesting phenomena, it is worthwhile to explore the effects of an attractive multi-body (three body) interaction in a spinor Bose gas. In this work, we shall study the effects of an attractive three body interaction on the SBHM to see the consequences on the phase diagrams, particularly the existence of odd-even asymmetry in the MI lobes.
\section{Model}
For spin-1 ultracold Bose gases loaded in an optical lattice, the Hamiltonian in presence an attractive three body interaction can be written as
\cite{PhysRevLett.81.742,ohmi,PhysRevLett.109.135302},
\begin{eqnarray}
H&=&-t\sum\limits_{<ij>}\sum\limits_{\sigma}(a^{\dagger}_{i\sigma}a_{j\sigma}+h.c)-\mu
\sum\limits_{i} n_{i}+U_{3}\sum\limits_{i}\delta_{n_{i},3}\nonumber \\ 
&+&\frac{U_{0}}{2}\sum\limits_{i}n_{i}(n_{i}-1)+\frac{U_{2}}{2}\sum\limits_{i}({\bf{S}}^{2}_{i}-2n_{i})
\label{bhm1chapt7}
\end{eqnarray}
Here operator $a^{\dagger}_{i\sigma}$ creates a boson at site $i$ with spin, $\sigma=\pm 1, 0$ and $t$ is the hopping amplitude from site $i$ to site $j$. The number operator is $n_{i}=\sum\nolimits_{\sigma}n_{i\sigma}$ where $n_{i\sigma}=a_{i\sigma}^{\dag}a_{i\sigma}$ and $\mu$ denotes the chemical potential. The total spin operator at a site $i$ is given by,
${\bf{S}}_{i}=a^{\dagger}_{i\sigma}{\bf{F}}_{\sigma\sigma'}a_{i\sigma'}$
where ${\bf{F}}_{\sigma\sigma'}$ are the components of spin-1 matrices. $U_{0}$ and $U_{2}$ refer to the two body spin independent and spin dependent on-site interaction potentials respectively while $U_{3}$ is the attractive three body interaction strength. It was experimentally found that for $^{23}Na$ atoms, the spin dependent interaction value is $U_{2}/U_{0}=0.0331$, refereed to antiferromagnetic (AF) while for $^{87}Rb$ atoms, it corresponds to -0.046, refereed to ferromagnetic interaction \cite{PhysRevLett.81.742}. 

In order to study the MI-SF phase transition related with Eq.(\ref{bhm1chapt7}), the hopping term is decoupled by employing the mean field approximation as\cite{PhysRevA.70.043628,PhysRevB.77.014503}
\begin{eqnarray}
a^{\dagger}_{i\sigma}a_{j\sigma} \simeq \langle
a^{\dagger}_{i\sigma} \rangle a_{j\sigma}
+a^{\dagger}_{i\sigma}\langle
a_{j\sigma}\rangle-\langle
a^{\dagger}_{i\sigma}\rangle \langle
a_{j\sigma}\rangle
\label{deco}
\end{eqnarray}
where the superfluid order parameter is defined as the equilibrium value of an
operator at a site $i$ as $\psi_{i\sigma}= \langle a_{i\sigma}\rangle$ assuming the order parameter is real. Using the above decoupling approximation, Eq.(\ref{bhm1chapt7}) can be written as the sum of the mean field Hamiltonians, $H=\sum\nolimits_{i}H^{MF}_{i}$ where $H^{MF}_{i}$ is given by,
\begin{eqnarray}
H^{MF}&=&\underbrace{\frac{U_{0}}{2}n(n-1)+\frac{U_{2}}{2}({\bf{S}}^{2}-2n)-\mu n+U_{3}\delta_{n,3}}_\text{$H^{0}$}\nonumber \\
&-&\underbrace{t\sum\limits_{\sigma}(\psi_{\sigma}a_{\sigma}+h.c)+t\sum\limits_{\sigma}\psi_{\sigma}^{2}}_\text{$H^{'}$}
\label{mf}
\end{eqnarray}
Here, $H^{0}$ and $H^{'}$ correspond to the unperturbed and perturbation Hamiltonians respectively and the site index, $i$ is removed due to homogeneity of the system. To obtain the equilibrium values of the SF order parameter, $\psi^{eq}$ and local densities, $\rho^{eq}$ in the ground state, the mean field Hamiltonian [Eq.(\ref{mf})] can be diagonalized by forming the matrix elements in the occupation number basis, $|n_{\sigma}\rangle$ as $\langle n_{\sigma}|H^{MF}|n'_{\sigma}\rangle$. Starting with an initial guess values for $\psi_{\sigma}$ and for $n=7$, we have continued the diagonalization process until the self consistent ground state, $\Psi^{g}$ is reached. Finally, we obtain the equilibrium SF order parameter and local densities as (dropping the subscript {\it{eq}} from hereafter)
\begin{eqnarray} 
\psi_{\sigma}&=&\langle\Psi^{g}|a_{\sigma}|\Psi^{g}\rangle\hspace{1cm} \psi=\sqrt{\psi^{2}_{+}+\psi^{2}_{0}+\psi^{2}_{-}}\nonumber\\ 
\rho_{\sigma}&=&\langle\Psi^{g}|n_{\sigma}|\Psi^{g}\rangle\hspace{1cm} \rho=\rho_{+}+\rho_{0}+\rho_{-}
\end{eqnarray}
\section{Results}
\subsection{MI lobes in the atomic limit}
\begin{figure}[!ht]
\centerline{ \hfill
\psfig{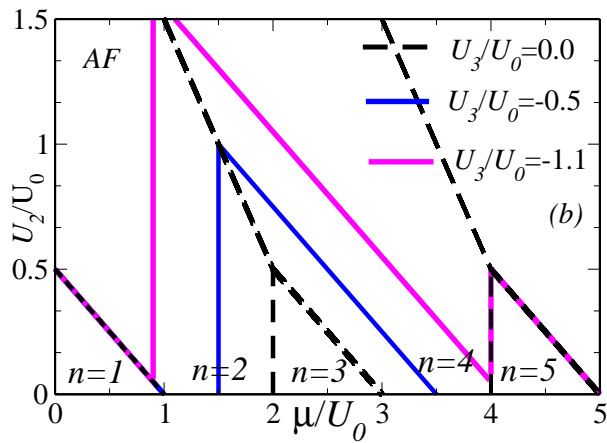}
   \hfill} 
 \caption{The width of the MI lobes in the atomic limit ($t=0$) with $U_{2}/U_{0}=0.05$ for different values of $U_{3}/U_{0}$. The plot shows that the odd-even asymmetry is destroyed around the third ($\rho=3$) MI lobe.}
\label{chapt71}
\end{figure}

In order to see how such an attractive three body interaction affects the MI-SF phase transition, let us first consider the atomic limit, that is, $t=0$ in Eq.(\ref{bhm1chapt7}). In the atomic limit, the system is completely in the insulating phase and hence the width of the MI lobe, ($\mu$) is the difference between the upper ($\mu_{+}$) and the lower ($\mu_{-}$) values of the chemical potential for a given occupancy, $n$ \cite{PhysRevA.83.013605}. For $t=0$, assuming a common eigenstate $|S,S_{z},n\rangle$ for $H^{0}$ where the corresponding operators, $S,S_{z},n$ commute with each other, the energy eigenvalue, $E^{0}(S,n)$ of $H^{0}$ is given by,
\begin{equation}
E^{0}(S,n)=\frac{U_{0}}{2}n(n-1)+\frac{U_{2}}{2}[S(S+1)-2n]-\mu n+U_{3}\delta_{n,3}
\end{equation} 
In the AF case, for the time being, we are assuming that the spin eigenvalue, $S=0$ for the even and $S=1$ for the odd MI lobes, as considered earlier without $U_{3}$ in Ref.\cite{PhysRevA.83.013605}. The boundaries of the MI lobe ($\mu_{\pm}$) can be obtained from the relation,
$E^{0}(S_{1},n-1)<E^{0}(S,n)<E^{0}(S_{2},n+1)$, where $S_{1,2}$ 
are the two spin eigenvalues corresponding to $n\mp 1$ occupancies. 
For the MI lobe with $n=3$, this inequality yields the following condition,
\begin{equation} 
2+\frac{U_{3}}{U_{0}}<\frac{\mu}{U_{0}}<3-2\frac{U_{2}}{U_{0}}-\frac{U_{3}}{U_{0}}
\end{equation}
Similarly for all other occupation densities, we have obtained the 
boundaries of each MI lobe separately and finally plotted them in Fig.\ref{chapt71} for 
different values of $U_{3}/U_{0}$. 
It shows that at a representative value for $U_{3}$, namely, $U_{3}/U_{0}
=-0.5$, although there is no change in the first and the fifth MI lobes but 
surprisingly the third MI lobe expands considerably, thereby engulfing the 
second ($\rho=2$) and the fourth ($\rho=4$) MI lobes. Also the critical value 
of $U_{2}/U_{0}$ at which all odd MI lobes disappear still remains at 0.5 \cite{PhysRevA.83.013605} 
for the 
first and the fifth MI lobes, while for the third MI lobe, it corresponds to 
$U^{c}_{2}>U_{0}/2-U_{3}$. Further, for a larger value of $U_{3}$, namely, $U_{3}/U_{0}=-1.1$, the 
third MI lobe grows compared to the other MI lobes by completely encroaching 
into the second and the fourth MI lobes. This implies 
that there exists a critical value of $U_{3}$ below which the second and the fourth MI lobes survive which is given by, $U^{c}_{3}/U_{0}=1+2U_{2}/U_{0}$. It is relevant to mention that for a scalar Bose gas, this critical value is of the order of $1$, that is $U_{3}\simeq -U_{0}$. While for a spinor Bose gas, this critical value depends on the two body spin dependent interaction, $U_{2}$. 
\subsection{MFA phase diagrams}
Now, to understand how such an attractive interaction modifies the odd-even asymmetry and hence the spin nematic-singlet formation in the MI phase, we shall
\begin{figure}[!h]
\centerline{ \hfill
\psfig{file=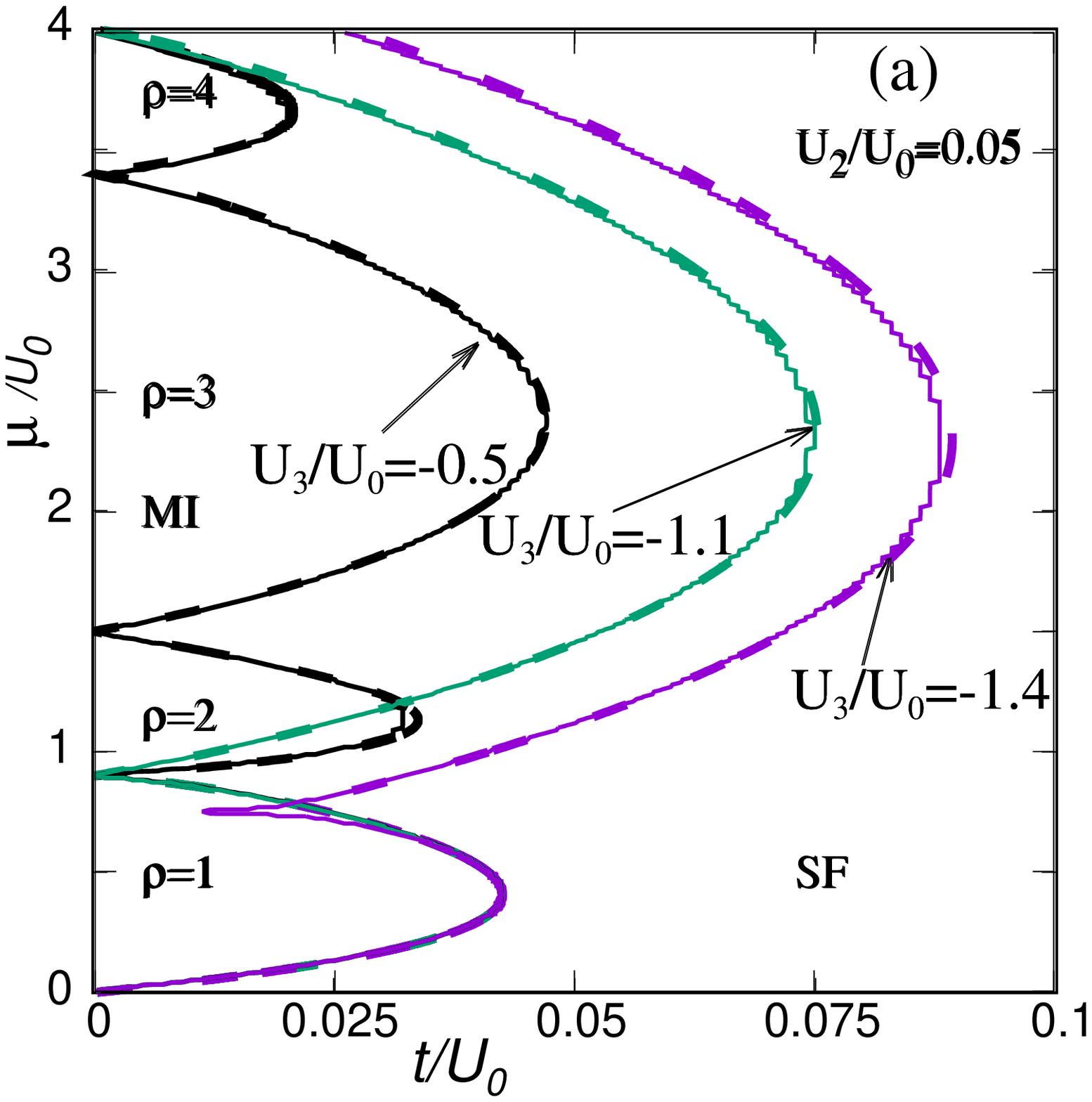,scale=0.45}
    \hfill}
\centerline{ \hfill 
     \psfig{file=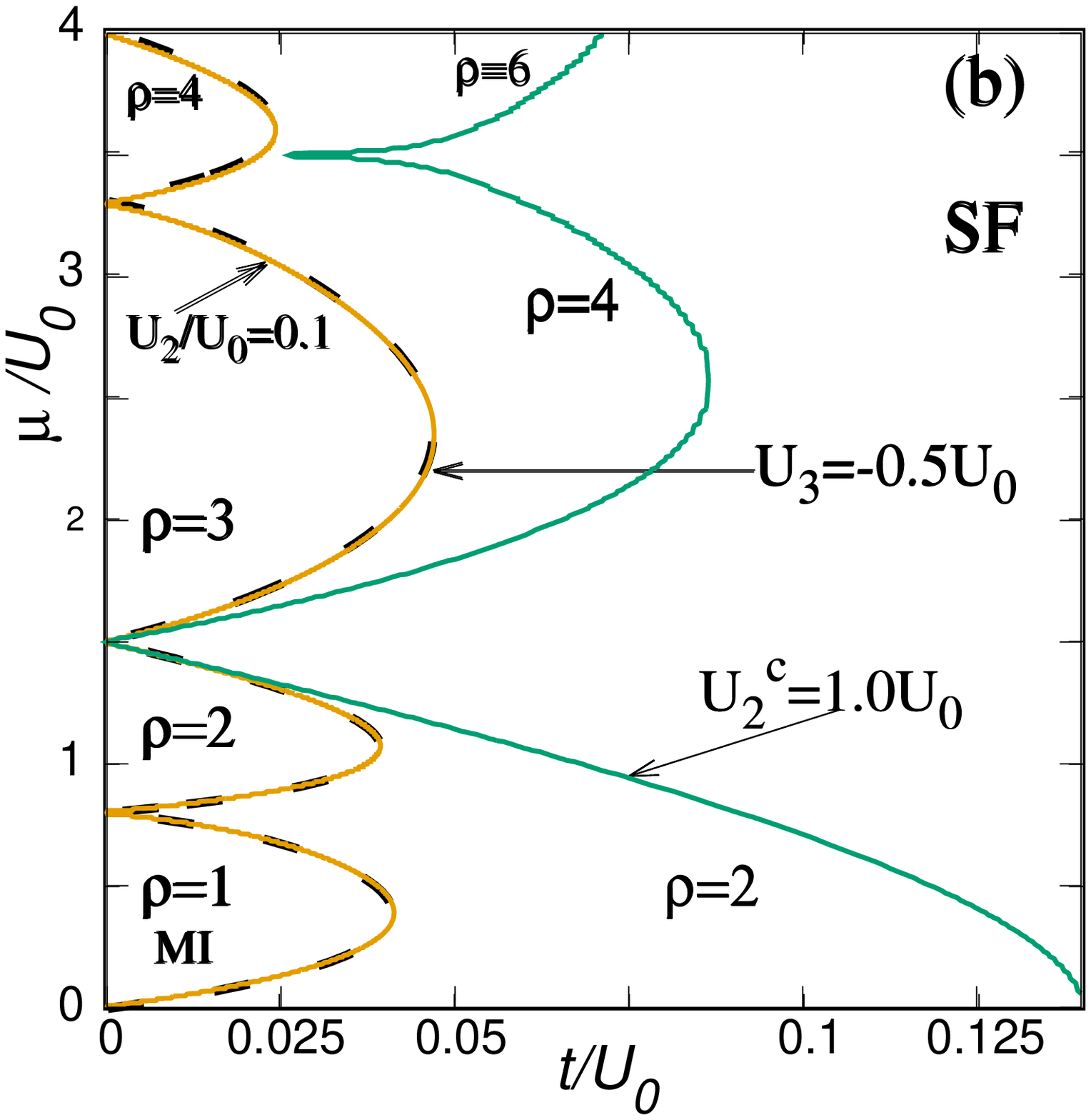,scale=0.45}    
         \hfill} 
 \caption{Phase diagrams from MFA (solid lines) and PMFA (dashed lines) for $U_{2}/U_{0}=0.05$ with attractive three body interaction potential, $U_{3}/U_{0}$ in (a). With increasing $U_{3}/U_{0}$, the asymmetry is destroyed around the third MI lobe. The phase diagrams corresponding to the higher values of spin dependent interaction, namely at $U_{2}/U_{0}=0.1$ and at a critical value, $U^{c}_{2}/U_{0}=1.0$ in (b).}
\label{chapt72}
\end{figure}
\begin{figure*}[!t]
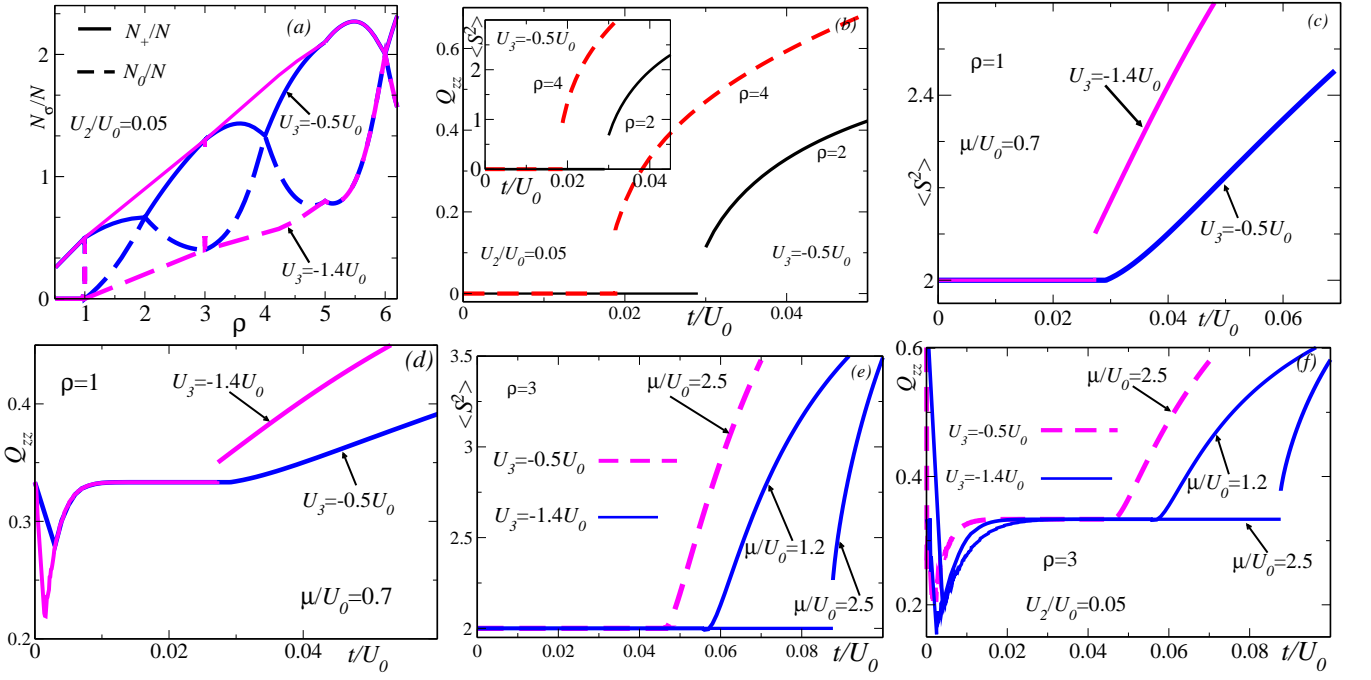

\centerline{ \hfill
\psfig{file=2a.eps,scale=0.23}
    \hfill
\psfig{file=2b.eps,scale=0.27}
    \hfill
\psfig{file=2d.eps,scale=0.27}}
  \centerline{  \hfill
     \psfig{file=2f.eps,scale=0.27}    
         \hfill
	\psfig{file=2e.eps,scale=0.27}
    \hfill
     \psfig{file=2g.eps,scale=0.27}    
         \hfill} 
 \caption{The population fraction, $N_{\sigma}/N$ with occupation densities, $\rho$ for $U_{2}/U_{0}=0.05$ with $U_{3}/U_{0}$ in (a). The total spin eigenvalue, $\langle S^{2}\rangle$ and the spin nematic order parameter, $Q_{zz}$ for $U_{2}/U_{0}=0.05$ corresponding to the even in (b) and the odd in (c)-(f) MI lobes.  The spin siglet formation is observed for the even MI lobes in (a). For the first ($\rho=1$) [(c), (d)] and the third ($\rho=3$) [(e), (f)] odd MI lobes with $U_{3}/U_{0}=-0.5$ and $U_{3}/U_{0}=-1.4$. For the third MI lobe, we have found a second order phase transition at $\mu/U_{0}=1.2$ and a first order at $\mu/U_{0}=2.5$ for $U_{3}/U_{0}=-1.4$.}
\label{chapt73}
\end{figure*} 
compute the mean field phase diagrams by turning on the hopping strength, $t/U_{0}$. The results are shown in Fig.\ref{chapt72} (a) and (b).

For $U_{2}/U_{0}=0.05$ [Fig.\ref{chapt72}(a)], we have found that at $U_{3}/U_{0}=-0.5$, the third MI lobe grows considerably by encroaching into both the second and the fourth MI lobes, while the first MI lobe remains unaffected as seen from Fig.\ref{chapt72}. At $U_{3}/U_{0}=-1.1$, the third MI lobe completely occupies the second and the fourth MI lobes, since the critical value of 
$U_{3}/U_{0}$ below which they exist corresponds to $1+2U_{2}/U_{0}$ as seen from 
Fig.\ref{chapt71}. 

Also, we have checked that at the critical value of $U^{c}_{2}/U_{0}=1.0$ for $U_{3}/U_{0}=-0.5$, the phase diagram only consists of all the even MI lobes, since all the odd MI lobes disappear at this critical value [Fig.\ref{chapt72}(b)]. At higher values of $U_{2}$, namely, $U_{2}/U_{0}=0.1$, the phase diagram shows similar effects as that of Fig.\ref{chapt72}(a) different values of $U_{3}/U_{0}$ [Fig.\ref{chapt72}(b)]. Increasing $U_{3}/U_{0}$ leads to the stabilization of the third MI lobe and hence the critical tunneling strength, $t_{c}/U_{0}$ also increases corresponding to the MI-SF phase transition.

The phase diagrams in Fig.\ref{chapt72}(a) and (b) show that, unlike the repulsive three body interaction \cite{NoorEPL2}, which retains the odd-even asymmetry in the MI lobes, here the attractive three body interaction diminishes the odd-even asymmetry around the second and the fourth MI lobes. This typically 
raises one concern, namely, whether will there any such asymmetry sustain beyond the fourth MI lobe.
To answer that, we have obtained the phase diagram corresponding to higher occupation densities and found that the chemical potential widths for the fifth ($\rho=5$) as well as the sixth ($\rho=6$) MI lobes remain unaltered for different values of $U_{3}/U_{0}$. This certainly underscores the importance of such interaction potential solely around the third MI lobe and also indicates that the asymmetry being intact beyond the fourth MI lobe. 
\subsection{Spin eigenvalue and spin nematic order parameter}
Further, in order to check if there is any phase transition involved with the SF phase, that is, from transverse polar (TP) to longitudinal polar (LP) state, we have shown the spin population fraction, $\rho_{\sigma}/ \rho$ as a function of the occupation densities, $\rho$ with $U_{3}/U_{0}$ for $U_{2}/U_{0}=0.05$ in Fig.\ref{chapt73} (a). It indicates that the hyperfine fractions corresponding to $\sigma=\pm 1$ are equal, that is, $\rho_{+}=\rho_{-}$. This is obvious since in absence of an external magnetic field, the total magnetization is conserved. Also no crossover is observed among the $\rho_{\pm}/\rho$ and $\rho_{0}/\rho$ components which suggests that the SF phase is completely in the TP state, as $\rho_{\pm}>\rho_{0}$. Further, at $U_{3}/U_{0}=-0.5$, we have found that the spin singlet formation corresponding to all even MI lobes occur where all the spin populations are equal, that is, $\rho_{\pm}/\rho=\rho_{0}/\rho$ \cite{PhysRevLett.102.140402}. Besides, it shows that the asymmetry is present around the fifth and the sixth MI lobes at larger values of $U_{3}/U_{0}$.
 
We have seen earlier that smaller values of attractive three body interaction show the spin singlet formation in the MI phase. It is therefore necessary to ascertain the order of phase transition as well as the spin eigenvalue variation from a spin singlet (nematic) MI phase to a SF phase. For $U_{3}/U_{0}=-0.5$, the spin eigenvalue, $\langle S^{2} \rangle=0$ for all the even MI lobes which indicates the formation of a spin singlet phase and it shows a first order phase transition to the SF phase Fig.\ref{chapt73} (b).
The spin eigenvalue, $\langle S^{2} \rangle$ corresponding to the first ($\rho=1$) is shown in Fig.\ref{chapt73} (c) and the third ($\rho=3$) in Fig.\ref{chapt73} (e) MI lobes with $U_{3}/U_{0}$ for $U_{2}/U_{0}=0.05$. At both values of $U_{3}$, the $\langle S^{2} \rangle=2$ indicates the existence of a spin nematic state corresponding to all odd MI lobes. Also, the phase transition from spin nematic MI to SF phase still continues to have a second order character for both the first ($\mu=0.7$) and third ($\mu=2.5$) MI phases at $U_{3}=-0.5U_{0}$. Interestingly, at larger value of $U_{3}$, namely, $U_{3}/U_{0}=-1.4$, the first MI lobe ($\mu=0.7$) shows a first order transition, while the third MI lobe displays a second order transition at $\mu=1.2$ and a first order transition at $\mu=2.5$ to the SF phase respectively [Fig.\ref{chapt73} (c) and (e)]. Further, we have computed the SF order parameter and the ground state energy. They show similar kind of phase transition as pointed out above corresponding to different MI lobes for both values of $U_{3}/U_{0}$.  

Now we shall look into the magnetic properties which signify the spin anisotropy of the spin singlet-nematic MI phases in presence of an attractive three body interaction. For that purpose, we have again calculated the $z$-component of the spin nematic order parameter which is defined as, $Q_{zz}=\langle {\bf{S}}^{2}_{z} \rangle-(1/3)\langle {\bf{S}}^{2} \rangle$ \cite{PhysRevLett.88.163001,PhysRevB.70.184434,PhysRevB.88.104509,PhysRevA.68.063602}. In the AF case, for the first and third MI lobes, $Q_{zz}$ is finite in the MI phase and it shows a second order transition at $U_{3}=-0.5U_{0}$, while a first as well as a second order phase transition for $U_{3}=-1.4U_{0}$ to the SF phase as pointed out earlier by us [Fig.\ref{chapt73} (d) and (f)]. For the spin singlet MI phase, the spin nematic order parameter vanishes since all the spin population fractions are equal, that is, $\rho_{\pm}/\rho=\rho_{0}/\rho$ and a first order transition to the SF phase is observed [Fig.\ref{chapt73} (b)]. 
\subsection{PMFA phase diagrams}
Since the spin eigenvalues, $S=0$ for the even and $S=1$ for the odd MI lobes in presence of three body interaction, we shall now focus on obtaining the analytic phase diagrams using perturbative mean field approaches (PMFA) to compare them with the numerical mean field phase diagram. Using $H^{'}$ as the perturbed Hamiltonian in Eq.(\ref{mf}), the ground state energy, $E_{n}$ which includes the first and second order corrections, is expressed 
in a series of $\psi$ \cite{PhysRevA.70.043628} as,
\begin{eqnarray}
E_{n}(\psi)&=&E^{0}+C_{2}(U_{0},U_{2},\mu,n,U_{3})\sum\limits_{\sigma}\psi^{2}_{\sigma}
\label{pert7}
\end{eqnarray}
The boundary between the MI-SF phase is obtained by putting $C_{2}(U_{0},U_{2},\mu,n,U_{3})=0$
due to minimization of $E_{n}(\psi)$ with respect to $\psi$. Using Eq.(\ref{pert7}), the boundary equation corresponding to the third MI lobe ($\rho=3$) is given by,
\begin{eqnarray}
\Bigg{(}\frac{1}{t}\Bigg{)}_{n=3}&=&\frac{8/15}{\mu-2U_{0}+3U_{2}-U_{3}}
+\frac{28/15}{3U_{0}+U_{2}-\mu-U_{3}}\nonumber
\\ &+&
\frac{5/3}{\mu-2U_{0}-U_{3}}+\frac{4/3}{3U_{0}-2U_{2}-\mu-U_{3}}
\label{oddchapt7}
\end{eqnarray}
Similarly for other MI lobes, we can obtain the boundary equations and the resultant phase diagrams are shown in Fig.\ref{chapt72} (a) and (b) (dashed lines).

At both values of $U_{2}/U_{0}$, the PMFA phase diagrams are in excellent agreement with the MFA results for all odd and even MI lobes corresponding to all values of $U_{3}/U_{0}$. However, we have checked that at smaller values of $U_{2}/U_{0}\le 0.03$, the MFA and the phase diagrams obtained via PMFA differ from each other slightly at the tip of the MI lobes, particularly it is prominent for the third MI lobe. Such deviation is due to inadequacy of the MFA to handle the fluctuations in reduced dimensions \cite{PhysRevA.70.043628,PhysRevLett.94.110403}. The above equation, which is quadratic in $\mu$, also shows that the critical tunneling strength, $t_{c}/U_{0}$ for the location of the MI-SF phase transition now increases as the three body interaction potential is enhanced.
\subsection{Phase diagram at $U_{3}=-1.4U_{0}$}
\begin{figure}[!ht]
\centerline{ \hfill
\psfig{file=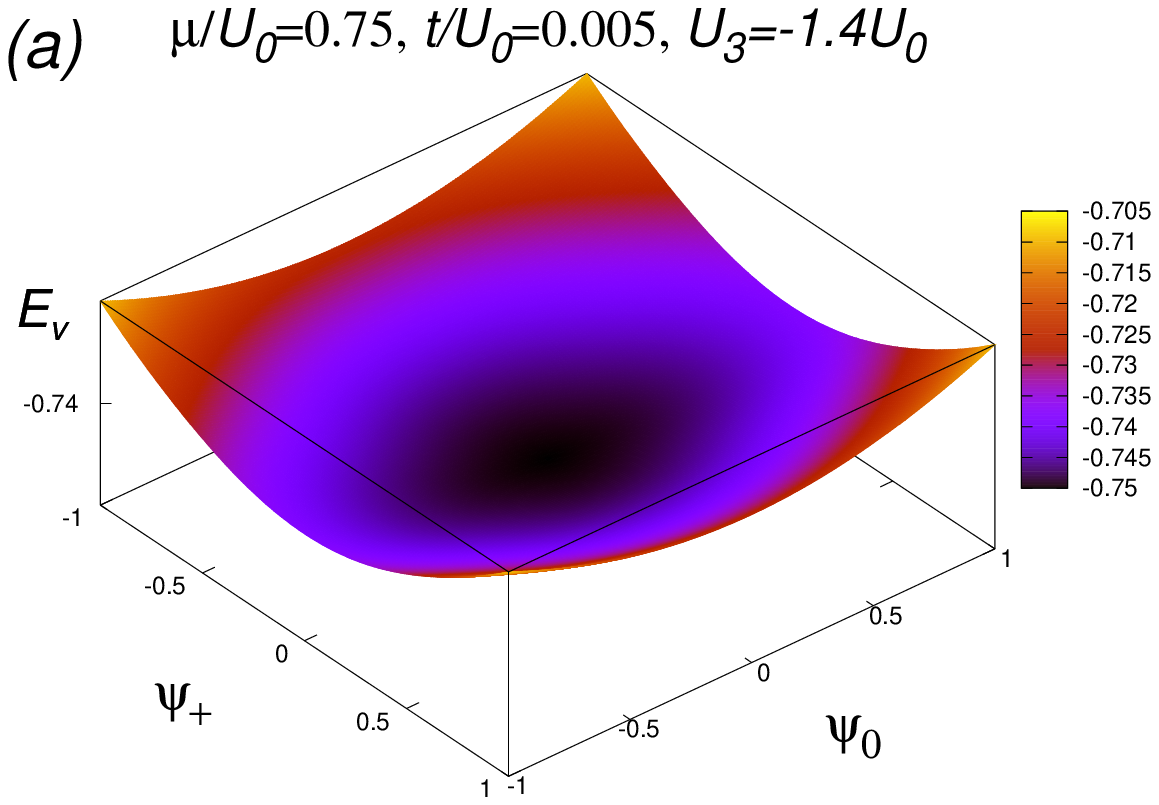,scale=0.33}
    \hfill\hfill 
     \psfig{file=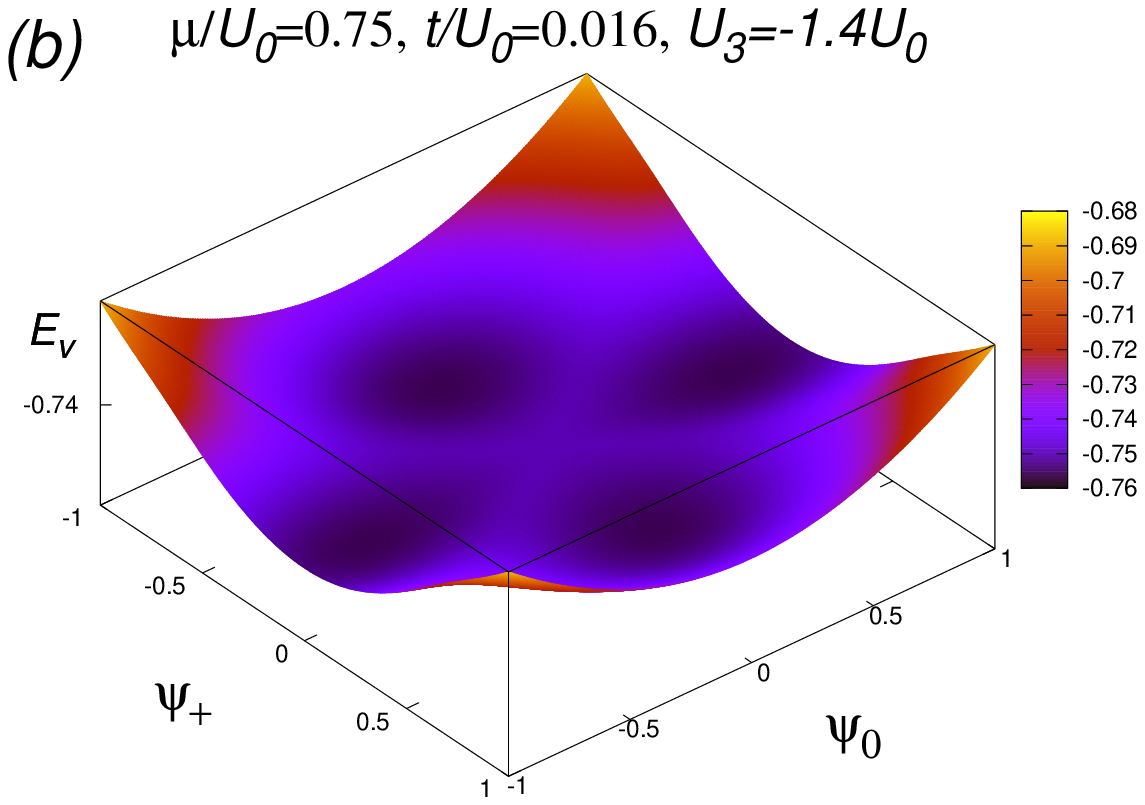,scale=0.33}}    
\centerline{\hfill
	\psfig{file=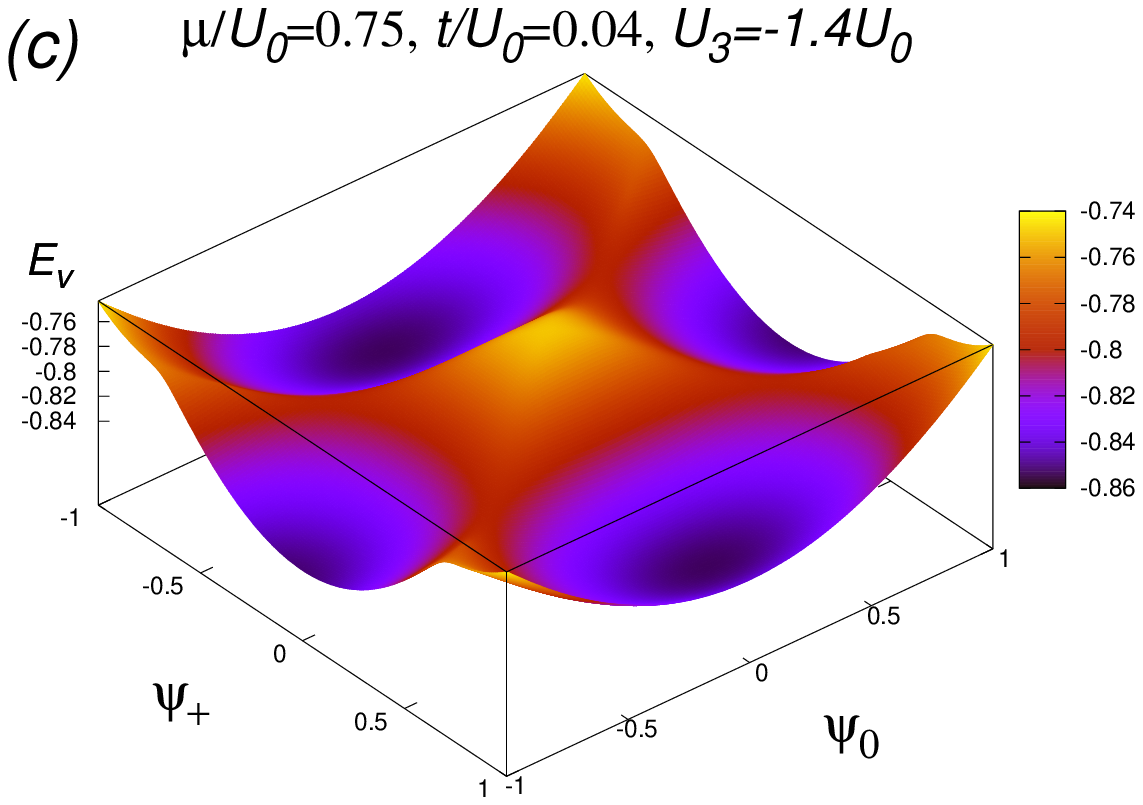,scale=0.33} 
	\hfill
	\psfig{file=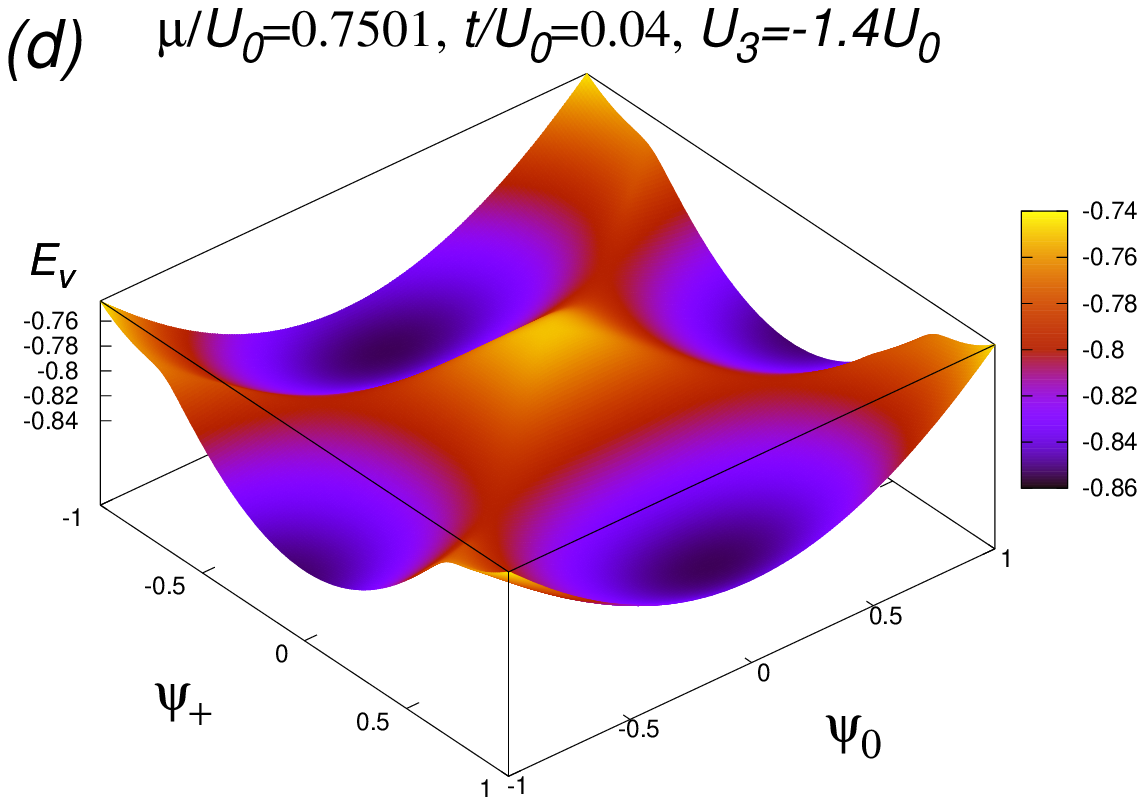,scale=0.33}\hfill} 
 \caption{The variational energy ($E_{v}$) behaviour for the first MI lobe ($\rho=1$) at $\mu/U_{0}=0.75$ in the MI phase at $zt/U_{0}=0.005$ in (a), near the MI-SF phase at $zt/U_{0}=0.016$ in (b) and finally in the SF phase at $zt/U_{0}=0.04$ in (c) for $U_{3}/U_{0}=-1.4$. $E_{v}$ shows a discontinuous transformation of the global minimum to four degenerate minima is a signature of first order transition for the first MI lobe ($\rho=1$). The third MI lobe ($\rho=3$) at $\mu/U_{0}=0.7501$ also shows first order transition from the MI to the SF phase in (d).}
\label{chapt76}
\end{figure}

In Fig.\ref{chapt73} (c)-(f), the spin eigenvalue and the spin nematic order parameters display different orders of phase transition at $U_{3}/U_{0}=-1.4$ which certainly necessitates a careful scrutiny of the complete phase diagram at different values of the chemical potential, $\mu$. For that purpose, we have studied the behaviour of variational energy, $E_{v}$ and the SF order parameter variation from the MI to the SF phase as a function of $t/U_{0}$ for both the $\rho=1$ and $\rho=3$ MI lobes with $\psi_{+}=\psi_{-}$ and $\psi_{0}$. For the first MI lobe ($\rho=1$) at $\mu/U_{0}=0.75$, the variational energy, $E_{v}$ shows a single minimum in the MI phase at $t/U_{0}=0.005$ [Fig.\ref{chapt76} (a)]. 
\begin{figure}[!ht]
\centerline{ \hfill
\psfig{file=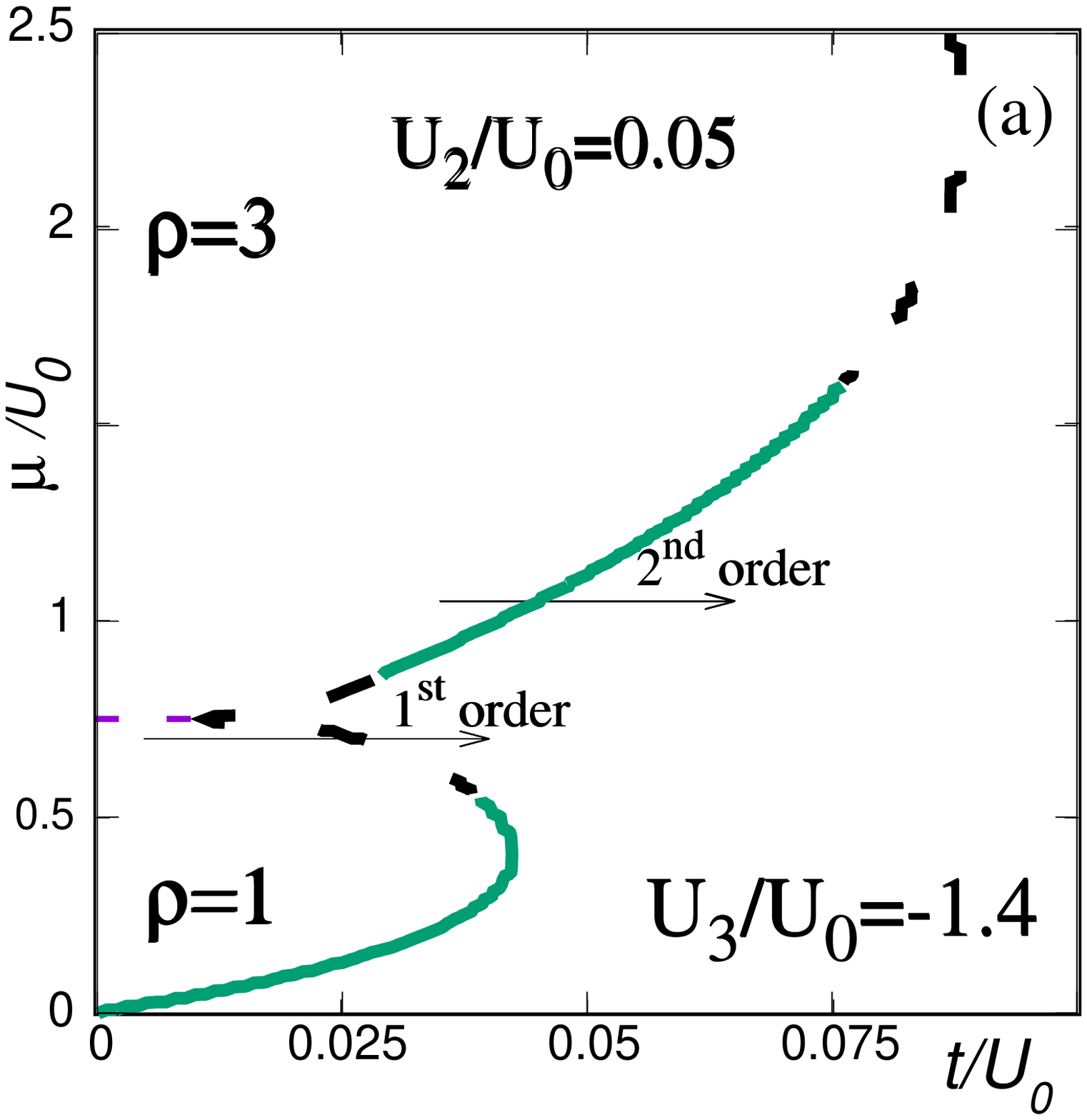,scale=0.26}
 \hfill
\psfig{file=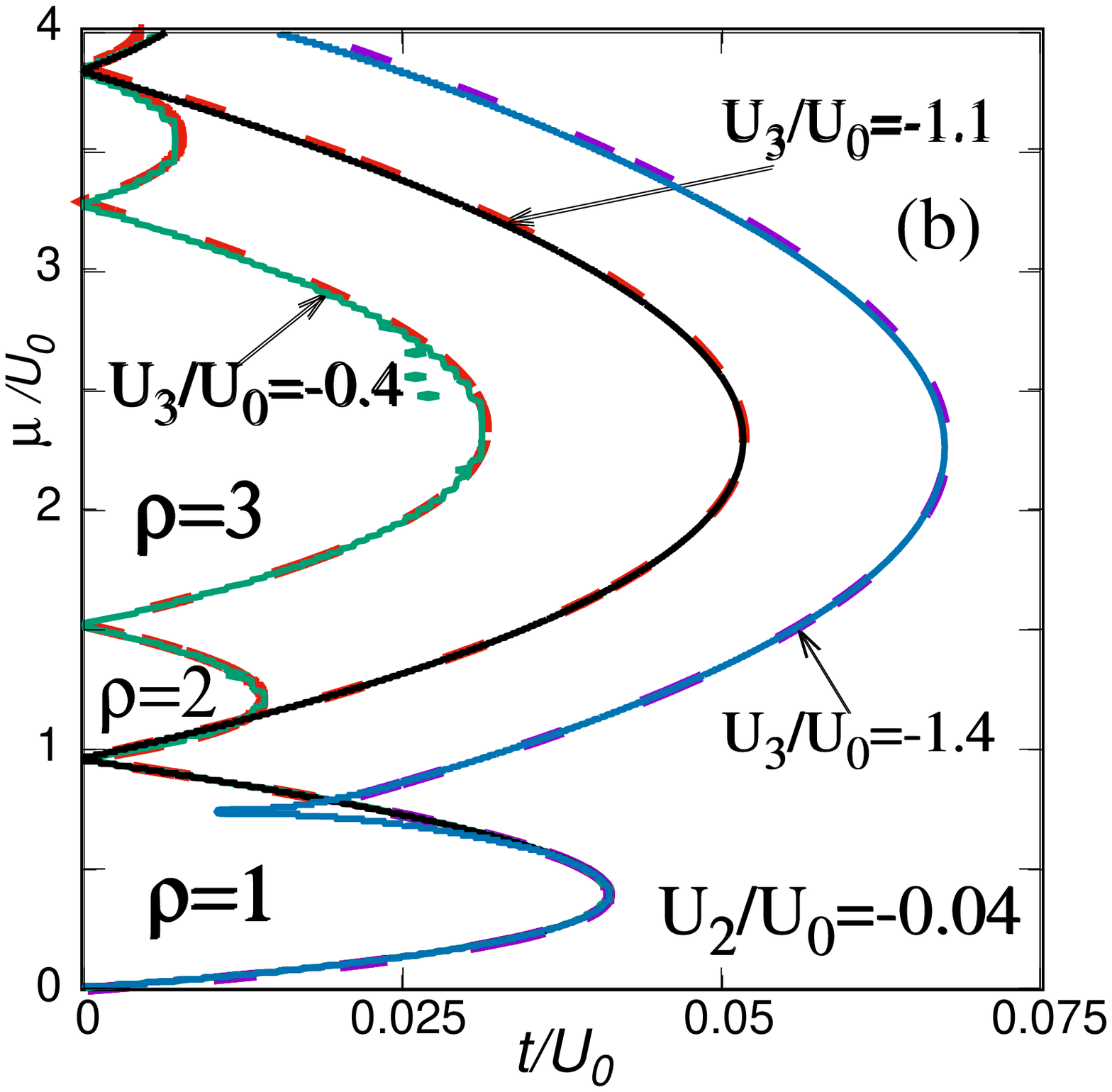,scale=0.26}
    \hfill} 
\caption{The detailed mean filed phase diagram depicting the first (dashed line) and second (solid lines) order transition for $U_{3}/U_{0}=-1.4$ with $U_{2}/U_{0}=0.05$ in (a). The MFA (solid lines) and PMFA (dashed lines) phase diagrams in the ferromagnetic case $U_{2}/U_{0}=-0.04$ with $U_{3}/U_{0}$ in (b).}
\label{chapt77}
\end{figure} 

As we approach toward the SF phase at $t/U_{0}=0.016$, four degenerate minima are formed [Fig.\ref{chapt76} (b)]. Finally at $t/U_{0}=0.005$ indicates a first order transition due to discontinuous evolution of the global minima [Fig.\ref{chapt76} (c)]. Similarly, $E_{v}$ for the third ($\rho=3$) MI lobe, that is, at $\mu/U_{0}=0.75+\epsilon$, where $\epsilon=0.001$, also demonstrates a first order phase transition at $t/U_{0}=0.04$ [Fig.\ref{chapt76} (d)]. On the other hand, the variational energy for the first MI lobe at $\mu/U_{0}=0.25$ shows a global minima in the MI phase and its continuous evolution in the SF phase signifies a second order transition. Further the order parameter and the variational energy indicate an analogous behaviour near the vicinity of the third ($\rho=3$) and the fifth ($\rho=5$) MI lobes. 

Thus the final phase diagram depicting the first and second order transition is shown in Fig.\ref{chapt77} (a) where the solid line represents a second order, while the dashed line indicates a first order phase transition. 

The phase diagrams corresponding to the ferromagnetic case ($U_{2}/U_{0}=-0.04$) for different values of $U_{3}/U_{0}$ are shown in Fig.\ref{chapt77} (b) and they are identical to a spin-0 Bose gas . We have also found similar kinds of phase transition corresponding to different MI lobes at $U_{3}=-1.4U_{0}$ which was pointed out earlier in Ref.\cite{PhysRevLett.109.135302}.
\section{Conclusion}
In this work, we have explored the consequences of an attractive three body interaction on the phase diagram of a SBHM using mean field and perturbative approaches. In the AF case, the third MI lobe expands by encroaching into the neighbouring second and fourth MI lobes which affects the odd-even asymmetry in the MI phase for smaller values of the three body interaction strength. However above the critical value of the three body strength, the third MI lobe predominantly occupies the whole phase diagram where it completely engulfs the neighbouring even MI lobes and thereby destroying such asymmetry around the second and the fourth MI lobes. Although the asymmetry remains intact beyond the fourth MI lobe at higher values of the potential strength.
\\\indent The spin eigenvalue and the nematic order parameter shows the formation of the spin singlet and nematic state corresponding to the even and odd MI lobes respectively for smaller values of the interaction strength. Also the order parameters show that the phase transition to the SF phase still maintains a first and second order character from the spin singlet and nematic MI phases respectively. 
At a considerable higher values of the three body strength, we have found a first as well as a second order transition at different location of the chemical potential corresponding to the first and the third MI lobes. This was confirmed by carefully scrutinizing the variational ground state energy and the spin eigenvalue and spin nematic order parameter. 
\\\indent The region in the vicinity of the first and third, as well as the third and the fifth MI lobes also show a first order phase transition which were subsequently confirmed from the behaviour of the variational energy and order parameters. Finally, all these MFA phase diagrams have splendid match with the analytical phase diagrams obtained using perturbative expansion (PMFA). Both the phase diagrams are in good agreement with each other however a small discrepancy is observed near the tip of the MI lobes for smaller values of the two body spin dependent interaction potential.
In the ferromagnetic case, the MFA phase diagrams are similar to that of the spin-0 Bose gas for different values of the three body interaction strength and they are in complete agreement with the analytical PMFA phase diagrams.
\bibliography{references} \bibliographystyle{aip}
\end{document}